# Deep learning-based auto-contouring of organs/structures-at-risk for pediatric upper abdominal radiotherapy

Mianyong Ding [1,2,3], Matteo Maspero [2,3], Annemieke S Littooij [1,2], Martine van Grotel [1], Raquel Davila Fajardo [1,2], Max M van Noesel [1,5], Marry M van den Heuvel-Eibrink [1,4], Geert O Janssens [1,2]

1. Princess Máxima Center for Pediatric Oncology, Utrecht, The Netherlands
2. Department of Radiation Oncology, Imaging and Cancer Division, University Medical Center Utrecht, Utrecht, The Netherlands
3. Computational Imaging Group for MR Diagnostics & Therapy, Center for Image Sciences, University Medical Center Utrecht, Utrecht, The Netherlands
4. Child Health Program, Wilhelmina Children's Hospital, University Medical Center Utrecht, Utrecht, The Netherlands
5. Division Imaging & Cancer, University Medical Center Utrecht, Utrecht, The Netherlands

# Abstract

**Purposes:**

This study aimed to develop a computed tomography (CT)-based multi-organ segmentation model for delineating organs-at-risk (OARs) in pediatric upper abdominal tumors and evaluate its robustness across multiple datasets.

**Materials and methods:**

In-house postoperative CTs from pediatric patients with renal tumors and neuroblastoma (n=189) and a public dataset (n=189) with CTs covering thoracoabdominal regions were used. Seventeen OARs were delineated: nine by clinicians (Type 1) and eight using TotalSegmentator (Type 2). Auto-segmentation models were trained using in-house (Model-PMC-UMCU) and a combined dataset of public data (Model-Combined).

Performance was assessed with Dice Similarity Coefficient (DSC), 95% Hausdorff Distance (HD95), and mean surface distance (MSD). Two clinicians rated clinical acceptability on a 5-point Likert scale across 15 patient contours. Model robustness was evaluated against sex, age, intravenous contrast, and tumor type.

**Results**

Model-PMC-UMCU achieved mean DSC values above 0.95 for five of nine OARs, while spleen and heart ranged between 0.90 and 0.95. The stomach-bowel and pancreas exhibited DSC values below 0.90. Model-Combined demonstrated improved robustness across both datasets. Clinical evaluation revealed good usability, with both clinicians rating six of nine Type 1 OARs above four and six of eight Type 2 OARs above three. Significant performance differences were only found across age groups in both datasets, specifically in the left lung and pancreas. The 0–2 age group showed the lowest performance.

**Conclusion**

A multi-organ segmentation model was developed, showcasing enhanced robustness when trained on combined datasets. This model is suitable for various OARs and can be applied to multiple datasets in clinical settings.

# 1 Introduction

Neuroblastomas and Wilms tumors are two of the most common pediatric abdominal solid tumors, representing 6% and 5% of childhood cancers, respectively [1]. Approximately 25% of Wilms tumor patients [2] and 64% of neuroblastoma patients [3] require radiotherapy in their treatment.

Radiotherapy relies on accurately delineating organs-at-risk (OARs) to minimize the dose to healthy tissues/organs. Typically, OAR delineation is performed on computed tomography (CT). The procedure is manual, time-consuming [4], and prone to inter-observer variability [5,6].

Recently, advancements in image processing and computational capacity, mainly through deep convolutional neural networks (CNNs) and dedicated hardware, have resulted in accurate and fast OAR auto-segmentation [7–9], leading to time savings during delineation [9], a reduction in inter-observer variability and differences in dose inconsistencies in clinical trials [10]. Numerous studies have focused on developing auto-segmentation models capable of delineating abdominal OARs from adult CTs [7,9,11,12], demonstrating good clinical acceptability for most OARs [12].

However, while significant advancements have been made in auto-segmentation on adults, advancements in children remain limited [13–15]. To our knowledge, a comprehensive model tailored for pediatric upper abdominal cases that meets clinical needs has yet to be developed. Furthermore, previous works have indicated that models trained exclusively on adult data tend to underperform on pediatric datasets, highlighting the need for models specifically developed using pediatric data [13,14].

Studies focused on pediatric abdominal imaging face several challenges, primarily the scarcity of comprehensive and reliable pediatric data. Only one public pediatric database exists, which includes over 20 manually delineated abdominal OARs [16]. Moreover, the incidence of pediatric abdominal tumors remains low [17], making it challenging to gather sufficient data from a single center. Data from different centers may exhibit considerable heterogeneity, including variations in patient demographics, imaging protocols, and scanner types. Models trained on datasets from a single center may have limited variability, resulting in poor generalization when applied to external datasets [18]. Conversely, using diverse and extensive datasets can improve auto-segmentation models' generalization capabilities [19]. Additionally, ongoing organ development in children and low imaging doses often used in pediatric patients further complicate the situation [16].

The primary goal of the current study is to develop a CT-based multi-organ model for delineating organs-at-risk in pediatric patients with upper abdominal tumors. Additionally, we study the model's robustness across a combined dataset of in-house and public data. Finally, model robustness was evaluated against sex, age, intravenous contrast (IV), and tumor type.

## 2 Materials and methods

### 2.1 Dataset description and preparation

This study investigates the performance of deep learning models on two datasets: (1) an in-house dataset from the Princess Máxima Center and UMC Utrecht (PMC-UMCU) and (2) a public dataset from Children's Wisconsin (PedSeg dataset) [16]. Both datasets contain pediatric CTs of thoracoabdominal regions, with patient demographics and imaging details summarized in Table 1 and age distribution provided in the Supplementary material (Figure S.1).

### 2.1.1 PMC-UMCU dataset

**Patient cohort and CT acquisition**: The dataset includes CTs performed between March 2015 and July 2023 at the UMCU for radiotherapy planning in a national cohort of pediatric patients with a renal tumor or neuroblastoma diagnosed at the PMC. All patients provided informed consent for the Princess Máxima Center Biobank procedure (Netherlands Trial Register NL7744, IRB approval number MEC-2016-739). After excluding eleven patients with missing annotations for more than four OARs, the PMC-UMCU dataset contained 80 patients with renal tumors and 102 patients with neuroblastoma, including seven patients with repeated imaging in the context of a second course of radiotherapy. In total, 189 postoperative CTs were included.

**OARs definitions**: We collected delineations for seventeen OARs: nine for clinician-made delineations and used for the radiotherapy planning (Type 1 OARs), while eight for delineations obtained with TotalSegmentator [7] (Type 2 OARs). Clinician-made contours were delineated by one of the two expert pediatric radiation oncologists in renal tumors or neuroblastoma [20]. Type 1 OARs include the spleen, lung (L+R), kidneys (L+R), heart, pancreas, and stomach-intestine-bowel (labeled "stomach-bowel"). For the heart, pancreas, and stomach-bowel, where more than 10% of cases had missing delineations (Figure S.2), TotalSegmentator was used to complement the missing delineations. For Type 2 OARs, all labels were generated by TotalSegmentator for the vertebrae, spinal cord (spinal canal), aorta abdominal, inferior vena cava, autochthonous muscles (L+R), and iliopsoas muscles (L+R) when clinical labels were unavailable.

**Management of overlapping structures**: Intersections between closely situated anatomical structures occasionally resulted in overlapping voxels. To manage these overlays, structures were in the following order: spleen, kidneys, heart > pancreas, liver > stomach-intestine-bowel >

lungs. Additionally, Type 1 OARs were prioritized over Type 2 due to their manual, clinician-based annotations, which were hypothesized of higher quality.

**Table 1.** Patient demographics, imaging acquisition, and image details.

| | Category | PMC-UMCU | PedSeg |
|---|---|---|---|
| Patient demographic | Cases | 182 | 189 |
| | Age (range) | 0-21 years (Median: 4.0) | 0-16 years (Median: 6.0) |
| | Gender ratio (M: F) | 100:82 | 101:88 |
| | Pathology | Renal tumor (80), neuroblastoma (102) | Not specified |
| Image acquisition and reconstruction | Machine | Philips Brilliance Big Bore (170), Siemens Biograph (19) | SOMATOM Definition AS+ (86), LightSpeed VCT (61), Revolution CT (42) |
| | IV | No (171) Yes (18) | No (11) Yes (178) |
| Images details | Number of CTs | 189 | 189 |
| | Number of slices (range) | 81-341 (Median: 183) | 114-354 (Median: 190) |
| | Voxel size[+] (mm) | 2.0 x 1.0 x 1.0 | 2.0 x 0.5 x 0.5 |
| | Image dimension[+] | 183, 512, 512 | 190, 512, 512 |
| | Tube voltage (kV) | 90-120 (Median: 90) | 70-120 (Median: 100) |
| | Exposure (mAs) | 31.0-1199.0 (Median 186.0) | 22.4-460.2 (Median:53.2) |

**Abbreviations:** + indicated as the median in the superior-inferior, left-right, and anterior-posterior directions; kV: kilo Volts; mAs: milli Ampere-seconds; CT: computed tomography; IV: intravenous contrast.

### 2.1.2 PedSeg dataset

The original dataset [16] contained 359 cases and was filtered to exclude patients with missing structures or slice counts outside the range of 80-400. From the 289 remaining cases, 189 CTs were randomly selected. Like the PMC-UMCU data, the annotations of the stomach, small intestine, and large intestine were combined into a single structure labeled "stomach-bowel." For Type 2 OARs, the spinal canal's original delineations remained, while delineations for other structures were generated using the TotalSegmentator.

The supplementary material (Figures S.3-S.6) includes examples of CTs and OAR delineations from the two datasets.

## 2.2 Model and evaluation metric

We used the automatic self-configuring framework nnUNet for model development [8], employing both 3D low-resolution (3D lowres) and 3D full-resolution (3D fullres) UNet architectures without ensembling strategies. Images were resampled to axial, sagittal, and coronal spacings: 3.12 mm, 1.59 mm, 1.59 mm (lowres) and 2.0 mm, 0.78 mm, 0.78 mm (fullres). Patch sizes were set to (axial, sagittal, coronal) 64x192x160 (fullres) and 80x160x160 voxels (lowres). Each model was trained for 1000 epochs, and the best-performing model was selected as the final model at the end of training. The default post-processing strategy of nnUNet was employed, using the validation set to decide whether to remove all but the largest connected component of each class. The hyperparameters used for all models are available on our GitHub repository (https://github.com/MMianyong/-PedAbdSeg-).

**Quantitative metrics**: Dice similarity coefficient (DSC), mean surface distance (MSD), and 95% Hausdorff distance (HD) were used to assess the segmentation accuracy for Type 1 organs. For the stomach-bowel structure, where intestinal regions were not wholly delineated, metrics were calculated solely for slices where the structure was present in the clinical contours, which we consider the ground truth. In renal tumor patients who had undergone nephrectomy, the false positive rate (FPR) was calculated to account for absent predicted kidneys. OARs without ground truth delineations were excluded from the evaluation.

**Qualitative metrics**: Clinical acceptability of model-generated contours was assessed using a 5-point Likert scale (Supplementary Material 4), where a score of four or above indicates that a contour is clinically acceptable with minor modifications, and a score of one indicates that it is unusable [21]. Auto-generated contours were manually reviewed and missing or poorly scored structures were manually controlled. In this study, we deemed structures with a mean score of three or higher across all cases sufficient to justify clinical use. Supplementary Material 4 depicts an example of the sheet presented to the reviewer.

## 2.3 Study design

The study consists of four parts (Figure 1): first, we built and evaluated a model using only PMC-UMCU data, then expanded it to a combined PMC-UMCU and PedSeg dataset. The best model was selected between the single dataset and combined dataset models for clinical evaluation. Finally, the model's robustness across different conditions was assessed.

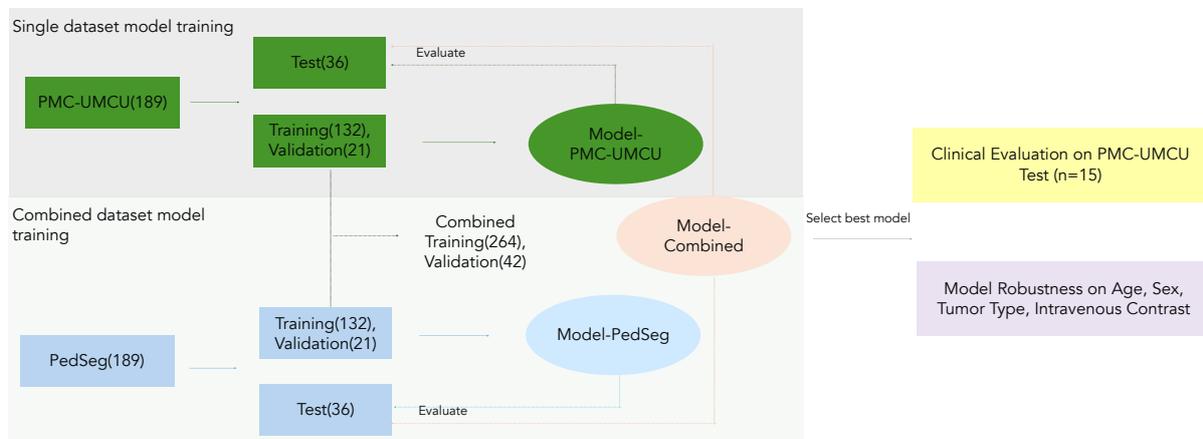

**Figure 1.** Workflow of the study design. The numbers in parentheses represent the number of CTs.

### 2.3.1 Single dataset model training

The PMC-UMCU dataset was chosen to identify the best auto-segmentation model architecture. The dataset was split at the patient level into representative subsets of 132 CTs for training, 21 for validation, and 36 for testing. The repeatability of the results is reported in Table S.1.

Two architectures, 3D lowres, and 3D fullres, were assessed using DSC, HD, and MSD on the test set. The best PMC-UMCU model was compared with TotalSegmentator on the same test set. The Wilcoxon signed-rank test was employed to compare the models' performance. Furthermore, the impacts of data size on model training were investigated, as reported in Table S. 2.

### 2.3.2 Combined dataset model training

The PedSeg dataset was split into 132:21:36 for training, validation, and testing, respectively. The sets were combined with the PMC-UMCU set to create a dataset with 264 CTs for training and 42 for validation. The best-performing architecture from 2.3.1 was trained on the combined dataset (Model-Combined) and PedSeg dataset (Model-PedSeg). Comparisons were made by evaluating predictions from Model-Combined, Model-PMC-UMCU, and Model-PedSeg on PMC-UMCU and PedSeg test data.

### 2.3.3 Clinical evaluation

The best-performing model from the evaluation from Section 2.3.2 was used for independent clinical validation. Model output from fifteen patients from the PMC-UMCU test set were randomly selected for review by two pediatric radiation oncologists (GJ, RDF). Each OAR, when present, was rated using a 5-point Likert scale, and the final mean score of each OAR was used to determine clinical usability. The inference time of the best-performing model was also reported.

### 2.3.4 Robustness of the model

The best-performing model was evaluated for robustness against sex (male vs. female), tumor type (renal vs. neuroblastoma), use of IV (with vs. without contrast), and age group. Patients were divided into four age groups for age-dependent analyses to ensure adequate sample sizes: 0-2, 3-4, 5-6, and ≥7. A 5-fold cross-validation on the combined 378 scans from PMC-UMCU and PedSeg, was performed to have a sufficiently large sample. Five different models were trained using a random split of 64:16:20 for training, validation, and testing. The Wilcoxon rank-sum test was applied to compare pairs of groups, except for the IV subgroup, due to its small sample size. A Bonferroni correction was applied, with corrected p-values < 0.05 indicating significant differences.

## 3 Results

### 3.1 Single dataset training

We developed and evaluated models on PMC-UMCU data using two architectures, 3D fullres and 3D lowres. Across the evaluation metrics for all OARs, no significant differences in performance were observed between the two architectures, with the 3D fullres resulting in the best average performances. Thanks to the highest DSC (↑) and lowest HD95 or MSD (↓) for most OARs (Table S.1), fullres was chosen as the preferred architecture.

The fullres model showed strong performance, with mean DSC values above 0.95 for the kidneys, liver, and lungs, and between 0.90 and 0.95 for the spleen and heart. The pancreas (0.70±0.11) and stomach-bowel (0.85±0.07) had mean DSC values below 0.90. The FPR for kidney predictions was 6/14 in 3D lowres and 3/14 in the 3D fullres.

Compared to TotalSegmentator, the 3D fullres model demonstrated higher mean DSC scores, with significant differences observed for all OARs (Figure S.8). Differences in mean DSC were around 0.01 for the liver and lungs, 0.01–0.1 for the spleen and pancreas, and 0.10–0.15 for the stomach-bowel and kidneys.

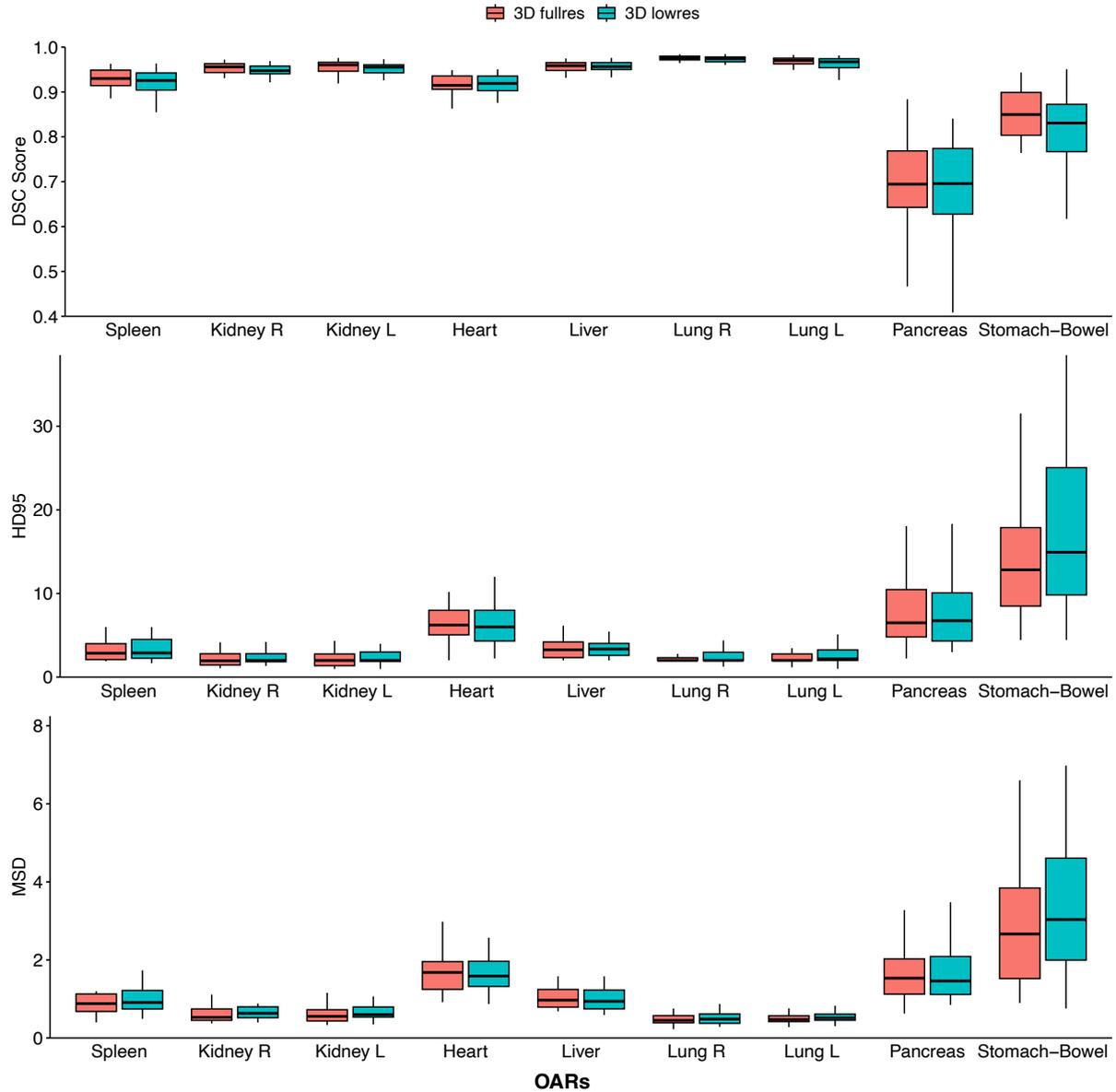

**Figure 2.** Box-whisker plot illustrates the performance distribution of the two best models of each configuration on the PMC-UMCU dataset using three metrics: DSC, HD95, and MSD. The horizontal line within each box represents the median, while the box spans the interquartile range (25th to 75th percentiles). The vertical lines (whiskers) extend from the boxes, and any points outside the whiskers represent outliers, which are not individually shown. No significant difference was found using the Wilcoxon signed-rank test.

The heart showed the largest difference (0.16). A case-by-case analysis of three cases with extremely low DSC (< 0.1) revealed that these cases belonged to very young age groups (0.7 years, 0.9 years, and 1.3 years). The FPR for the removed kidney was 2/14.

## 3.2 Combined dataset training

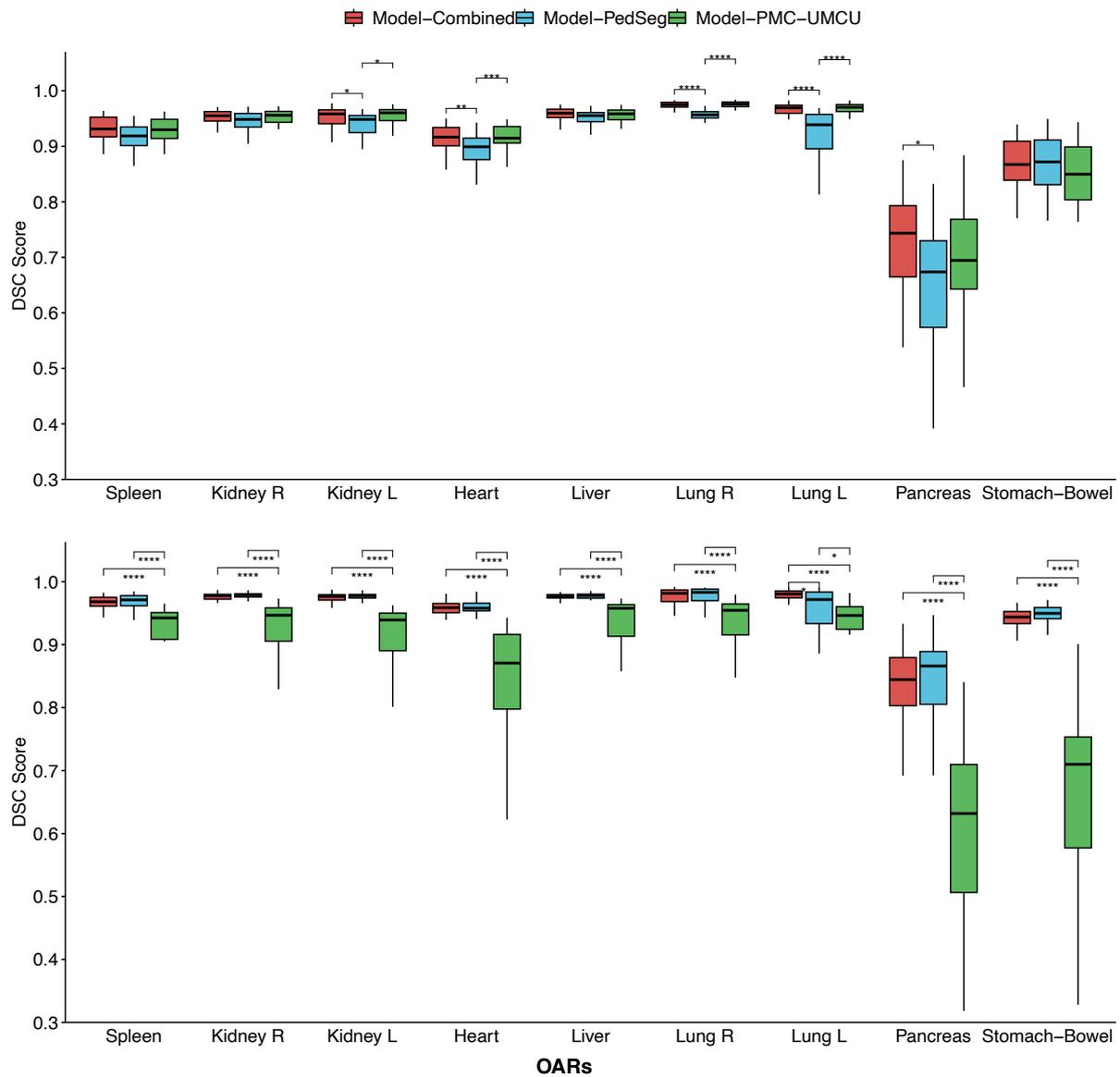

**Figure 3.** The box-whisker plot shows DSC scores of Model-Combined, Model-PedSeg, and Model-PMC-UMCU on PMC-UMCU (top) and PedSeg (bottom) datasets. Model-Combined: 3D fullres model based on the combined dataset. Model-PMC-UMCU: 3D fullres model solely based on the PMC-UMCU dataset. Model-PedSeg: 3D fullres model based on the PedSeg dataset. Annotations above the boxes indicate the significance of the Wilcoxon signed-rank test, with **** for $p \leq 0.0001$, *** for $0.0001 < p \leq 0.001$, ** for $0.001 < p \leq 0.01$, and * for $p \leq 0.05$.

To compare models trained on data from combined datasets versus uncombined datasets, we found that models trained on a single dataset generally performed well on their dataset but were less robust on other datasets. For instance, Model-PMC-UMCU performed significantly worse across all OARs on PedSeg, while Model-PedSeg showed the poorest performance for the left kidney, heart, and lungs on PMC-UMCU.

In contrast, Model-Combined demonstrated greater robustness across both datasets (Figure 3). Its performance was comparable to models trained on datasets, with no significant differences and all differences in mean DSC within 0.04 (Table S.3). The only significant difference was observed for the left lung on PedSeg, where Model-Combined showed a 0.04 improvement in mean DSC. Additionally, in PMC-UMCU nephrectomy cases, the FPR decreased to 2/14.

## 3.3 Clinical evaluation

Model-Combined was selected for clinical evaluation due to its better robustness. The average inference time, including post-processing, was 64.5 seconds per patient, using an Intel Xeon Gold 6240 CPU at 2.60 GHz, 251 GB of RAM, and a Tesla V100 GPU with 32 GB. Nine of the fifteen patients evaluated had neuroblastoma, and six had renal tumors with only one remnant kidney. Prediction examples are shown in Figure 4, S.9, and S.10.

For Type 1 OARs, all organs—except the heart, pancreas, and stomach-bowel—received average scores above four from both raters, indicating suitability for clinical use (Table S.4). The heart and pancreas were rated above three, with the largest disagreement for the pancreas (3.87±0.74 vs. 5±0). The stomach-bowel scored (3.07±0.26 vs. 2.80±0.86) near the borderline of clinical usability.

For Type 2 OARs, both raters agreed the spinal canal scored above four, and the autochthon(L+R), iliopsoas muscles(L+R), and vertebrae scored above three. Vena cava inferior was rated below two and deemed unusable. There was disagreement on the aorta-abdominal (1.47± 0.94 vs. 3.00±1.61). Despite these differences, all Type 1 OARs except for the stomach-bowel and Type 2 OARs except for the vena cava inferior and aorta-abdominal received scores above three, demonstrating clinical utility for most OARs.

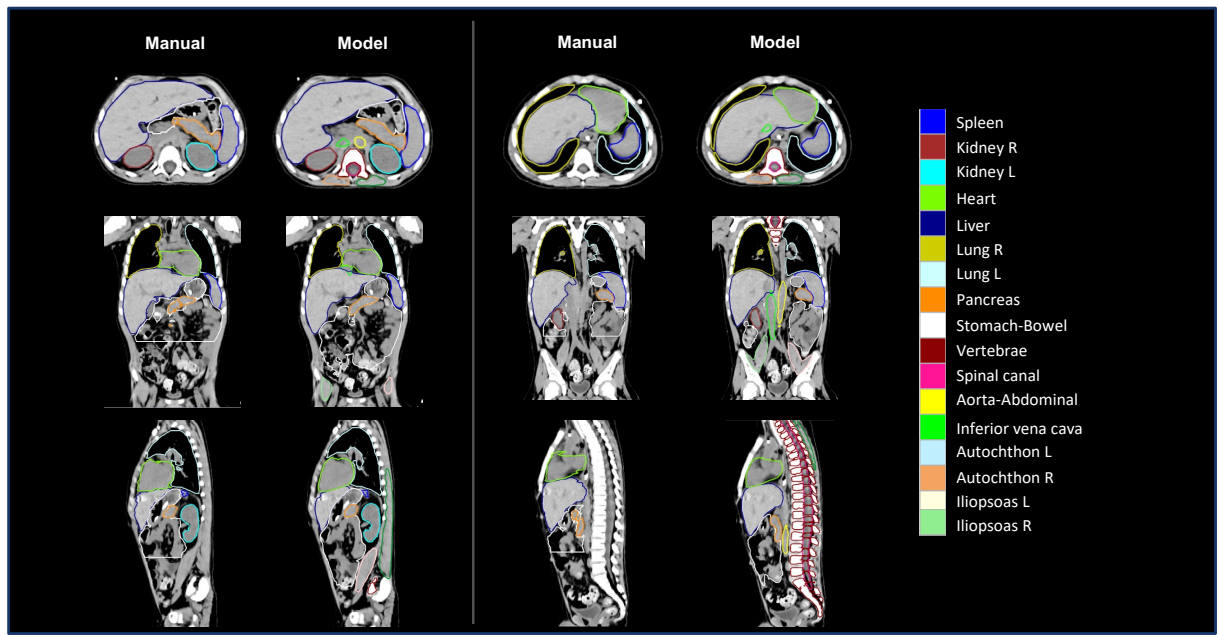

**Figure 4.** Example contours from multiple planes of a single patient showing type 1 OARs from manual contours and both type 1 and type 2 OARs generated by the Model-Combined.

## 3.4 Model robustness

The model showed no significant differences in most OARs by sex (in both PMC-UMCU and PedSeg) or tumor types (in PMC-UMCU) (Figures S13-S15). Using IV resulted in higher mean DSC for most OARs at both datasets (Figure S.16, S.17). Age-related variations were observed, with distinct patterns between the two datasets. In PMC-UMCU, significant age-dependent differences were found only in the left lung and pancreas (Figure 5), while in PedSeg, significant age-related differences were observed across all OARs except the bowel-intestine (Figure S.12). These findings suggest that the model's performance may vary with patient age, an important consideration for its clinical use. For the left lung, both datasets showed increasing mean DSC with age, with significant differences between younger and older groups. The pancreas showed increasing performance with age in PedSeg, while in PMC-UMCU, it had the highest mean DSC in the 3-4 age group. In both cohorts, significant differences were observed between the 0-2 age and older groups, with the 0-2 group consistently showing lower DSCs.

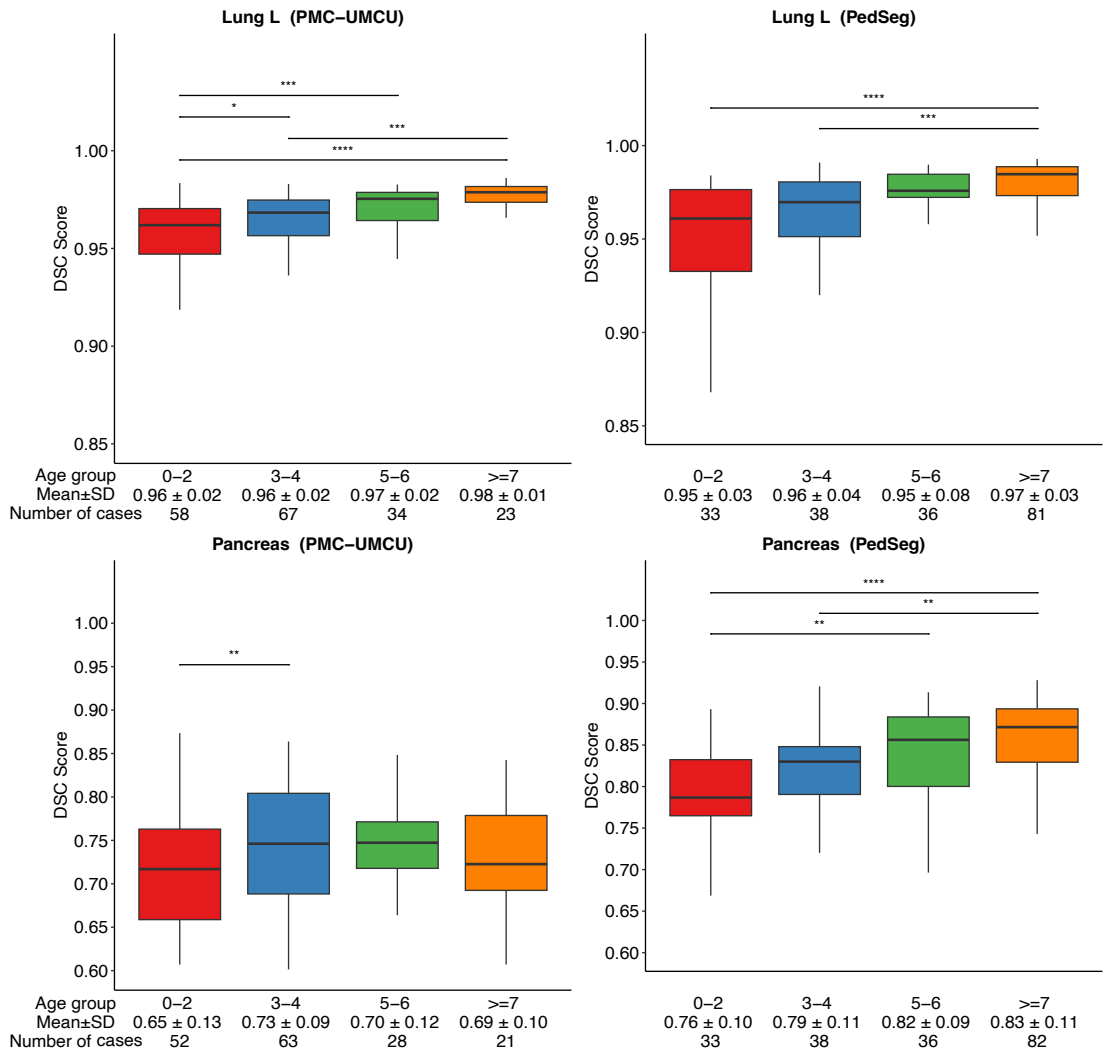

improved performance when combined with a public pediatric dataset. Clinical evaluation confirmed that most type 1 OARs were clinically acceptable, and most type 2 OARs were clinically usable.

It is necessary to develop pediatric-specific models: even when a model is trained in a small dataset (n=10), it outperformed TotalSegmentator in our in-house pediatric cohort (Table S.2). Few studies have investigated auto-segmentation for pediatric upper abdominal OARs, and none have developed or shared a single model capable of segmenting all OARs specifically for upper abdominal target volumes applicable for pediatric radiotherapy [13–15].

Comparing the mean DSCs reported in other pediatric studies for type 1 OARs (excluding the stomach-bowel), our model's performance aligns with studies using deep learning-based models on PedSeg and internal datasets [13–15]. Pancreas segmentation remains challenging across studies, with DSCs ranging from 0.73 to 0.79 [13–15]. Our model achieved a DSC of 0.73 on internal testing and 0.81 on PedSeg. The lower performance may be due to the pancreas's low contrast with surrounding structures, the lack of peri-pancreatic fatty tissue in young children, and large anatomical variability.

One of the main challenges in our research was using real-world data from clinical practice contours without modification. In our in-house dataset, we encountered missing labels or incomplete delineations for some patients. Differences in organ definitions between datasets or between the clinic and TotalSegmentator, e.g., for the stomach-bowel, posed further challenges. Despite these limitations, we successfully developed a model resulting in clinically acceptable delineations for most type 1 OARs. This suggests that even without extensive review, clinical contours can still help develop models intended for clinical use. For type 2 OARs, where annotations were unavailable, TotalSegmentator helped generate labels, resulting in 6/8 clinically usable structures. However, some OARs, such as the aorta and inferior vena cava, remained challenging. OARs that did not achieve optimal clinical evaluation may require additional manual correction. Incorporating active learning could improve the efficiency of contour corrections [22].

Pediatric datasets are rare, and our results indicate that a sample size of about 50 cases is sufficient to build the model in our in-house dataset. A further increase in sample size did not improve performance. Training on data from two heterogeneous cohorts improved the model's robustness without significantly reducing performance at either dataset compared to individually trained models. Generally, data privacy remains a barrier to sharing patient information across institutions. Federated learning, which enables model training without direct data sharing, may offer a promising solution for developing more robust segmentation models[23]. We will explore federated learning to strengthen the model's robustness in future work.

The model's robustness was evaluated across patient subgroups, with age-dependent variations observed. PMC-UMCU showed fewer OARs with age-related performance differences, possibly due to the distribution of patients skewing towards younger ages compared to PedSeg. Across all age groups, the 0–2 age group remained the most challenging, with significantly lower DSCs than older groups, likely due to less sample size, greater anatomical variability, and difficulties in imaging younger patients. Despite using a stratified validation approach, limited sample sizes may have reduced the statistical power of the tests, particularly for the patients older than seven years in the PMC-UMCU) cohort. Similarly, the small sample size of groups with or without IV in two datasets prevented statistical testing.

Our models demonstrate good clinical usability based on PMC-UMCU data, making, on average, predictions in less than 1 minute compared to 1 to 2 hours estimated for manual delineation for a single patient. We will hold a workshop to evaluate the model and the time needed to revise its output among multiple observers. Our model is shared online and can be applied to other centers for pediatric abdominal radiotherapy, promising to facilitate radiotherapy planning.

# 5 Conclusion

A multi-organ segmentation model for deep learning-based auto-contouring of OARs in pediatric upper abdominal radiotherapy was developed using data from a single and combined dataset. Incorporating an additional dataset improved the model's robustness. Clinical evaluation revealed high performance for most OARs. The model is publicly available, facilitating the adoption of automatic OAR delineation in pediatric radiotherapy.

# Declaration of Generative AI and AI-assisted technologies in the writing process

While preparing this work, the author (M. Ding) used ChatGPT-4 for grammar and spelling revision. After using this tool/service, the author(s) reviewed and edited the content as needed and take(s) full responsibility for the publication's content.

# Acknowledgments

This project is supported by the HE/MSCA co-fund project no. 101081481, and by the KiTZ-Maxima Twinning Program, stimulating collaborations and joint projects between the Hopp Children's Cancer Center (KiTZ) in Heidelberg and the Princess Máxima Center in Utrecht.

# Author Contribution

**Mianyong Ding**: Conceptualization, Data curation, Formal analysis, Investigation, Methodology, Software, Validation, Visualization, Roles/Writing - original draft, Writing - review & editing.

**Matteo Maspero**: Conceptualization, Data curation, Formal analysis, Funding acquisition, Methodology, Resources, Software, Supervision, Validation, Roles/Writing - original draft, Writing - review & editing.

**Annemieke S Littooij**: Writing - review & editing.

**Martine van Grotel**: Writing - review & editing.

**Raquel Davila Fajardo:** Formal analysis, Writing - review & editing.

**Max M van Noesel**: Writing - review & editing.

**Marry M van den Heuvel-Eibrink**: Funding acquisition, Project administration, Resources, Supervision, Writing - review & editing.

**Geert O Janssens**: Conceptualization, Data curation, Formal analysis, Funding acquisition, Project administration, Resources, Supervision, Writing - review & editing.

# 6 Reference

[1] L.L. de Faria, C.P. Clementino, F.A.S. E Véras, D. da C. Khalil, D.Y. Otto, M.O. Filho, L. Suzuki, M.A. Bedoya, Staging and Restaging Pediatric Abdominal and Pelvic Tumors: A Practical Guide, Radiographics 44 (2024). https://doi.org/10.1088/1361-6560/ac678a.

[2] M.M. Van Den Heuvel-Eibrink, J.A. Hol, K. Pritchard-Jones, H. Van Tinteren, R. Furtwängler, A.C. Verschuur, G.M. Vujanic, I. Leuschner, J. Brok, C. Rübe, A.M. Smets, G.O. Janssens, J. Godzinski, G.L. Ramírez-Villar, B. De Camargo, H. Segers, P. Collini, M. Gessler, C. Bergeron, F. Spreafico, N. Graf, Rationale for the treatment of Wilms tumour in the UMBRELLA SIOP–RTSG 2016 protocol, Nature Reviews Urology 2017 14:12 14 (2017) 743–752. https://doi.org/10.1038/nrurol.2017.163.

[3] A. Fukushima, V. Hande, K. Wakeham, M.B. Barton, M.S. Zaghloul, D.C. Moreira, N. Bhakta, K. Pritchard-Jones, M. Sullivan, B. Mazhar Qureshi, P.N. Njiraini, A. Polo, Estimation of the optimal radiotherapy utilization rate for childhood neuroblastoma, Radiotherapy and Oncology 197 (2024) 110343. https://doi.org/10.1016/J.RADONC.2024.110343.

[4] W. Chen, C. Wang, W. Zhan, Y. Jia, F. Ruan, L. Qiu, S. Yang, Y. Li, A comparative study of auto-contouring softwares in delineation of organs at risk in lung cancer and rectal cancer, Scientific Reports 2021 11:1 11 (2021) 1–8. https://doi.org/10.1038/s41598-021-02330-y.

[5] R. McCall, G. MacLennan, M. Taylor, N. Lenards, B.E. Nelms, M. Koshy, J. Lemons, A. Hunzeker, Anatomical contouring variability in thoracic organs at risk, Medical Dosimetry 41 (2016) 344–350. https://doi.org/10.1016/J.MEDDOS.2016.08.004.

[6] C.L. Brouwer, R.J.H.M. Steenbakkers, E. van den Heuvel, J.C. Duppen, A. Navran, H.P. Bijl, O. Chouvalova, F.R. Burlage, H. Meertens, J.A. Langendijk, A.A. van 't Veld, 3D Variation in delineation of head and neck organs at risk, Radiation Oncology 7 (2012) 1–10. https://doi.org/10.1186/1748-717X-7-32/FIGURES/4.

[7] J. Wasserthal, H.C. Breit, M.T. Meyer, M. Pradella, D. Hinck, A.W. Sauter, T. Heye, D.T. Boll, J. Cyriac, S. Yang, M. Bach, M. Segeroth, TotalSegmentator: Robust Segmentation of 104 Anatomic Structures in CT Images, Radiol Artif Intell 5 (2023). https://doi.org/10.1148/RYAI.230024.

[8] F. Isensee, P.F. Jaeger, S.A.A. Kohl, J. Petersen, K.H. Maier-Hein, nnU-Net: a self-configuring method for deep learning-based biomedical image segmentation, Nat Methods 18 (2021) 203–211. https://doi.org/10.1038/s41592-020-01008-z.

[9] X. Chen, S. Sun, N. Bai, K. Han, Q. Liu, S. Yao, H. Tang, C. Zhang, Z. Lu, Q. Huang, G. Zhao, Y. Xu, T. Chen, X. Xie, Y. Liu, A deep learning-based auto-segmentation system for organs-at-risk on whole-body computed tomography images for radiation therapy, Radiotherapy and Oncology 160 (2021) 175–184. https://doi.org/10.1016/J.RADONC.2021.04.019.

[10] M. Thor, A. Apte, R. Haq, A. Iyer, E. LoCastro, J.O. Deasy, Using Auto-Segmentation to Reduce Contouring and Dose Inconsistency in Clinical Trials: The Simulated Impact on RTOG 0617, International Journal of Radiation Oncology*Biology*Physics 109 (2021) 1619–1626. https://doi.org/10.1016/J.IJROBP.2020.11.011.

[11] H. Kim, J. Jung, J. Kim, B. Cho, J. Kwak, J.Y. Jang, S. wook Lee, J.G. Lee, S.M. Yoon, Abdominal multi-organ auto-segmentation using 3D-patch-based deep convolutional neural network, Sci Rep 10 (2020). https://doi.org/10.1038/S41598-020-63285-0.

[12] C. Yu, C.P. Anakwenze, Y. Zhao, R.M. Martin, E.B. Ludmir, J. S.Niedzielski, A. Qureshi, P. Das, E.B. Holliday, A.C. Raldow, C.M. Nguyen, R.P. Mumme, T.J. Netherton, D.J. Rhee, S.S. Gay, J. Yang, L.E. Court, C.E. Cardenas, Multi-organ segmentation of abdominal structures from non-contrast and contrast enhanced CT images, Sci Rep 12 (2022) 19093. https://doi.org/10.1038/S41598-022-21206-3.

[13] K. Kumar, A.U. Yeo, L. McIntosh, T. Kron, G. Wheeler, R.D. Franich, Deep Learning Auto-Segmentation Network for Pediatric Computed Tomography Data Sets: Can We Extrapolate From Adults?, International Journal of Radiation Oncology*Biology*Physics (2024). https://doi.org/10.1016/J.IJROBP.2024.01.201.

[14] E. Somasundaram, Z. Taylor, V. V. Alves, L. Qiu, B. Fortson, N. Mahalingam, J. Dudley, H. Li, S.L. Brady, A.T. Trout, J.R. Dillman, Deep-Learning Models for Abdominal CT Organ Segmentation in Children: Development and Validation in Internal and Heterogeneous Public Datasets, American Journal of Roentgenology (2024). https://doi.org/10.2214/AJR.24.30931/SUPPL_FILE/24_30931_SUPPL.PDF.

[15] P.M. Adamson, V. Bhattbhatt, S. Principi, S. Beriwal, L.S. Strain, M. Offe, A.S. Wang, N.J. Vo, T. Gilat Schmidt, P. Jordan, Technical note: Evaluation of a V-Net autosegmentation algorithm for pediatric CT scans: Performance, generalizability, and application to patient-specific CT dosimetry, Med Phys 49 (2022) 2342–2354. https://doi.org/10.1002/MP.15521.

[16] P. Jordan, P.M. Adamson, V. Bhattbhatt, S. Beriwal, S. Shen, O. Radermecker, S. Bose, L.S. Strain, M. Offe, D. Fraley, S. Principi, D.H. Ye, A.S. Wang, J. van Heteren, N.J. Vo, T.G. Schmidt, Pediatric chest-abdomen-pelvis and abdomen-pelvis CT images with expert organ contours, Med Phys 49 (2022) 3523–3528. https://doi.org/10.1002/MP.15485.

[17] K. Nakata, M. Colombet, C.A. Stiller, K. Pritchard-Jones, E. Steliarova-Foucher, Incidence of childhood renal tumours: An international population-based study, Int J Cancer 147 (2020) 3313–3327. https://doi.org/10.1002/IJC.33147.

[18] A. Barragan-Montero, A. Bibal, M.H. Dastarac, C. Draguet, G. Valdes, D. Nguyen, S. Willems, L. Vandewinckele, M. Holmstrom, F. Lofman, K. Souris, E. Sterpin, J.A. Lee, Towards a safe and efficient clinical implementation of machine learning in radiation oncology by exploring model interpretability, explainability and data-model dependency, Phys Med Biol 67 (2022) 11TR01. https://doi.org/10.1088/1361-6560/AC678A.

[19] J. Hofmanninger, F. Prayer, J. Pan, S. Röhrich, H. Prosch, G. Langs, Automatic lung segmentation in routine imaging is primarily a data diversity problem, not a methodology problem, Eur Radiol Exp 4 (2020) 1–13. https://doi.org/10.1186/S41747-020-00173-2/FIGURES/5.

[20] G.O. Janssens, P. Melchior, J. Mul, D. Saunders, S. Bolle, A.L. Cameron, L. Claude, K. Gurtner, K.P. van de Ven, M. van Grotel, S. Harrabi, Y. Lassen-Ramshad, N. Lavan, H. Magelssen, X. Muracciole, T. Boterberg, H. Mandeville, J. Godzinski, N. Graf, M.M. van den Heuvel-Eibrink, C. Rübe, The SIOP-Renal Tumour Study Group consensus statement on flank target volume delineation for highly conformal radiotherapy, Lancet Child Adolesc Health 4 (2020) 846–852. https://doi.org/10.1016/S2352-4642(20)30183-8.

[21] A.C. Kraus, Z. Iqbal, R.A. Cardan, R.A. Popple, D.N. Stanley, S. Shen, J.A. Pogue, X. Wu, K. Lee, S. Marcrom, C.E. Cardenas, Prospective Evaluation of Automated Contouring for CT-Based Brachytherapy for Gynecologic Malignancies, Adv Radiat Oncol 9 (2024) 101417. https://doi.org/10.1016/j.adro.2023.101417.

[22] C. Qu, T. Zhang, H. Qiao, J. Liu, Y. Tang, A.L. Yuille, Z. Zhou, AbdomenAtlas-8K: Annotating 8,000 CT Volumes for Multi-Organ Segmentation in Three Weeks, Adv Neural Inf Process Syst 36 (2023). https://arxiv.org/abs/2305.09666v2 (accessed September 15, 2024).

[23] Z. Zheng, Y. Hayashi, M. Oda, T. Kitasaka, K. Misawa, K. Mori, Federated 3D multi-organ segmentation with partially labeled and unlabeled data, Int J Comput Assist Radiol Surg (2024) 1–14. https://doi.org/10.1007/S11548-024-03139-6/FIGURES/7.